\documentclass[10pt,twocolumn,letterpaper]{article}

\usepackage[pagenumbers]{cvpr} 

\usepackage{graphicx}
\usepackage{amsmath}
\usepackage{amssymb}
\usepackage{booktabs}
\usepackage{multirow,tabularx}

\usepackage[pagebackref,breaklinks,colorlinks]{hyperref}

\usepackage[capitalize]{cleveref}
\crefname{section}{Sec.}{Secs.}
\Crefname{section}{Section}{Sections}
\Crefname{table}{Table}{Tables}
\crefname{table}{Tab.}{Tabs.}

\def\cvprPaperID{5994} 
\def\confName{CVPR}
\def\confYear{2023}

\begin{document}

\title{Structural Similarity: When to Use Deep Generative Models on Imbalanced Image Dataset Augmentation}

\author{Chenqi Guo$^{1}$~~~~ Fabian Benitez-Quiroz$^{1}$~~~~ Qianli Feng$^{1,2}$~~~~ Aleix Martinez$^{1,2}$\\
    $^1$The Ohio State University~~~~$^2$Amazon~~~~\\
    {\tt\small \{guo.1648, benitez-quiroz.1, feng.559, martinez.158\}@osu.edu}
    }
\maketitle

\begin{abstract}
   Improving the performance on an imbalanced training set is one of the main challenges in nowadays Machine Learning. One way to augment and thus re-balance the image dataset is through existing deep generative models, like class-conditional Generative Adversarial Networks (cGAN) or Diffusion Models by synthesizing images on each of the tail-class. Our experiments on imbalanced image dataset classification show that, the validation accuracy improvement with such re-balancing method is related to the image similarity between different classes. Thus, to quantify this image dataset class similarity, we propose a measurement called Super-Sub Class Structural Similarity (SSIM-supSubCls) based on Structural Similarity (SSIM). A deep generative model data augmentation classification (GM-augCls) pipeline is also provided to verify this metric correlates with the accuracy enhancement. We further quantify the relationship between them, discovering that the accuracy improvement decays exponentially with respect to SSIM-supSubCls values.
\end{abstract}

\section{Introduction}
\label{sec:intro}

\begin{figure}
\centering
\includegraphics[width=1.0\linewidth]{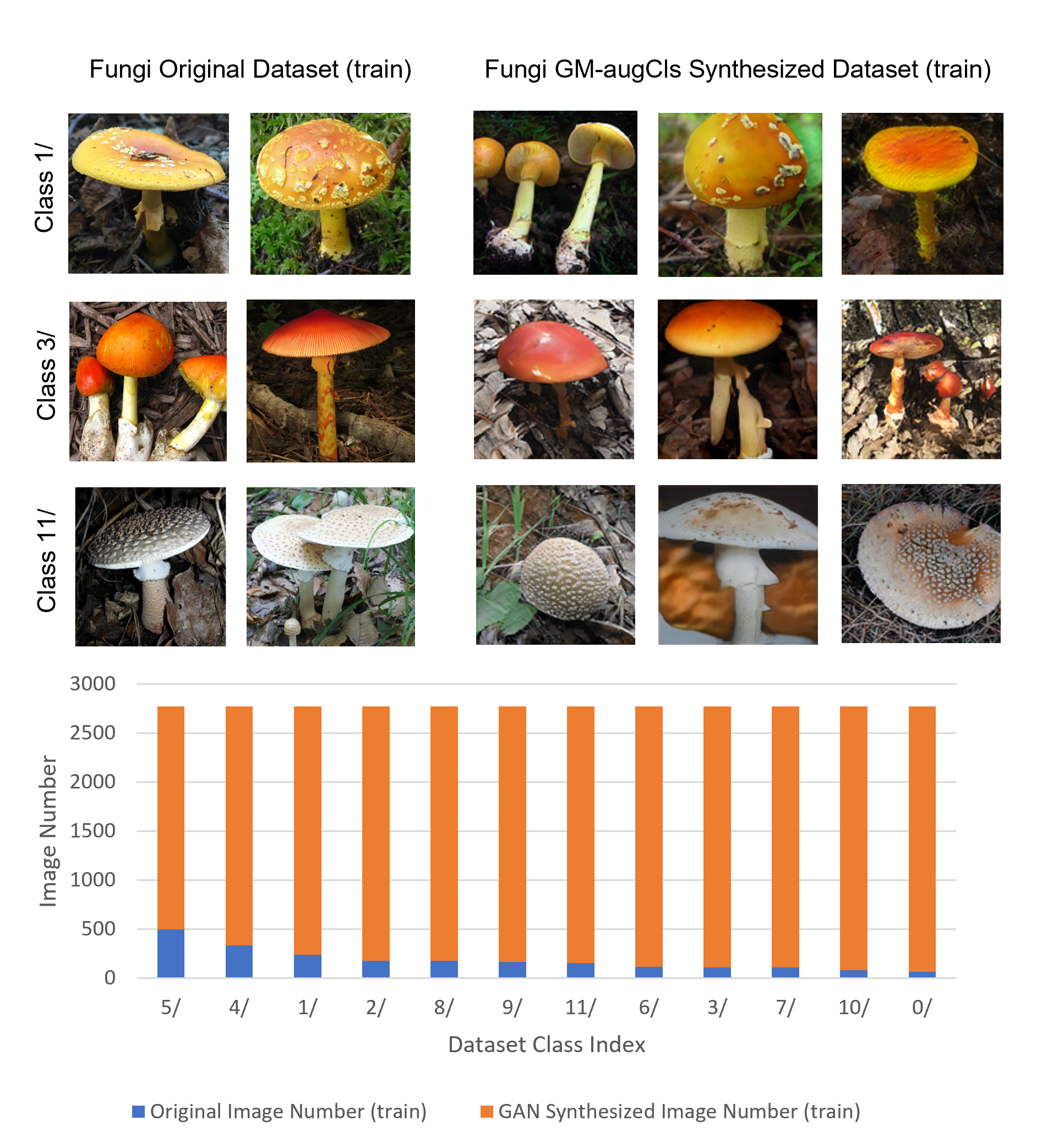}
\caption{An example of data augmentation application with our proposed deep generative model data augmentation classification (GM-augCls) pipeline on iNaturalist-2019 Fungi dataset. ({\it Top}) exampled original and synthesized images for visualization. ({\it Bottom}) bar plot of the image number in each class for the original and synthesized dataset. After the augmentation procedure, the new dataset becomes balanced. In this paper, a class similarity metric for image dataset, called Super-Sub Class Structural Similarity (SSIM-supSubCls) is constructed for measuring the image similarity between different classes. We further quantify the relationship between this metric value and the classification accuracy improvement after GM-augCls procedure, discovering that the accuracy improvement decays exponentially with respect to SSIM-supSubCls values.
}
\label{fig:teaser}
\end{figure}

Machine learning with imbalanced dataset is still a challenging and open task. Many efforts have been made to mitigate the bias towards the majority classes and thus enhance the generalization of classifiers trained on such image datasets. Common strategies concerning adding weights for each class in the loss function \cite{cui2019classbalancedloss}, re-balancing the dataset by over-sampling the minority classes or under-sampling the majority classes \cite{BrancoTR15} are widely used. As one of the most popular over-sampling techniques, geometric transformations like rotations or mirroring may interfere with orientation-related features in the image. SMOTE \cite{SMOTE2002} can be used to solve the over-sampling problem by creating synthetic data based on the nearest neighbors of a sample in an embedding or image space. Other techniques attempt to use images from the web \cite{openlongtailrecognition}, knowledge transfer \cite{junya2021supercharging} and GAN architectures generating artificial images \cite{tanaka2019data,mariani2018bagan,antoniou2018data,utkarsh2020elastic} to enlarge the minority classes. 
For image related tasks, one way to augment the dataset is through deep generative models, like class-conditional Generative Adversarial Networks (cGAN) or Diffusion Models, by synthesizing artificial images on the tail-classes to make the training set balanced.

\begin{figure}
\centering
\includegraphics[width=1.0\linewidth]{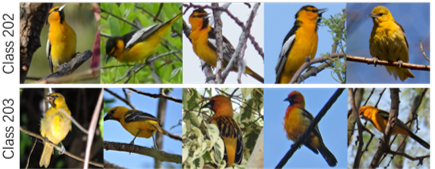}
\caption{An example of iNaturalist-2019 Birds dataset with high image similarity between classes. Here class 202 and class 203 share similar background, bird shape and color. Hence, expert specialities are required to distinguish between these two species.
}
\label{fig:similar_subCls}
\end{figure}

However, some classification tasks are inherently harder than others, especially in the absence of data. For instance, intuitively the classification task in Figure \ref{fig:teaser} is easier than Figure \ref{fig:similar_subCls}: the two classes in Figure \ref{fig:similar_subCls} share similar background, object shape and color, and the distinction between these two bird species requires subject matter expertise. In this case, deep generative models would produce unfaithful reconstructions, since small but important differences between class object attributes are inherently difficult to be generated. Therefore, the downstream classification performance may not be improved and in some scenarios even negatively impacted. Hence the question comes naturally when we can use deep generative model data augmentation to refine classification performance.

In our work, we propose a metric measuring the dataset image similarity between different classes based on Structural Similarity (SSIM), called Super-Sub Class Structural Similarity (SSIM-supSubCls). 
For such metric, higher value suggests higher class similarity. Furthermore, we provide a pipeline named deep generative model data augmentation classification (GM-augCls), which uses StyleGAN2 based cGAN \cite{shahbazi2022collapse} or guided-diffusion model \cite{guidedDiffusion} to generate images. The synthesized images are then passed through a classifier trained on the original dataset to calculate the likelihood of each new image being part of the target class. Only images with a high likelihood are appended to the training set. Experiments reveal that the accuracy improvement decays exponentially with respect to the metric values.
The whole idea is shown in image Figure \ref{fig:teaser}.

We summarize our contributions as follows:
\begin{itemize}
  \item We propose a metric measuring image dataset class similarity 
  (section \ref{sec:SSIM-supSubClsMetric} SSIM-supSubCls). 
  A lower metric value implies lower image similarity across different classes. 
  \item We propose a data augmentation pipeline re-balancing the imbalanced image dataset using existing deep generative models, like cGAN or diffusion models, jointly with a classifier as image filtering technique. 
  \item We evaluate our proposed metric and pipeline on imbalanced dataset, including iNaturalist-2019 \cite{Horn2018CVPR}, flowers \cite{flowersDatasetKaggle}, UTKFace \cite{UTKFaceDatasetKaggle} and scene \cite{sceneDatasetKaggle} dataset. Qualitative and quantitative results suggest that the pipeline can faithfully generate high quality images. Experiments also show that the top1 validation accuracy increment after dataset re-balancing decays exponentially with respect to the metric values. With this exponential curve fitted, one can easily predict the expected accuracy improvement for any image dataset from its SSIM-supSubCls value.
\end{itemize}

\section{Related Works}
\label{sec:formatting}

\subsection{Imbalanced Dataset}
Many real-world datasets are not uniformly distributed over different classes. Such imbalanced datasets bring in biases towards the majority classes and affect the generalizability of classifiers trained on them. Widely used solutions include imposing inductive bias, or statistical re-sampling or re-weighting to alleviate the extreme imbalance. To enhance the learning performance of imbalanced data, \cite{jiang2021improving} sought a principled approach to select unlabeled data in the wild scaling up the original dataset using contrastive learning approach. 
Showing that real world dataset often has a long-tailed and open-ended distribution, \cite{openlongtailrecognition} defined Open Long-Tailed Recognition (OLTR) as learning from such naturally distributed data and optimized the classification accuracy by handling tail recognition robustness with dynamic meta-embedding over a balanced test set. To deal with the shift in the testing label distribution caused by training set imbalance, \cite{junjiao2020posterior} derived a post-training prior re-balancing technique through a KL-divergence based optimization. 

\subsection{Data Augmentation and Deep Generative Models}
When the training image dataset is imbalanced, the classification accuracy can be significantly deteriorated, especially for the tail classes. One way to mitigate this problem is through data augmentation on minority classes. Traditional methods include geometric transformation like image rotations or translations, re-sampling with SMOTE \cite{SMOTE2002} or ADASYN \cite{ADASYN4633969}. Efforts have also been made to use deep generative models, like class-conditional GAN (cGAN) \cite{shahbazi2022collapse} or diffusion models \cite{guidedDiffusion} to restore the dataset balance by synthesizing minority-class samples. 

\cite{tanaka2019data} experimented with different GAN architectures generating new training data for classification tasks and showed that the GAN improved the results compared with the original imbalanced dataset. \cite{mariani2018bagan} proposed a balancing generative adversarial network (BAGAN) architecture to generate new images for minority-class. \cite{antoniou2018data} used a Data Augmentation Generative Adversarial Network (DAGAN) for data augmentation by taking data from a source domain and generalising it to a different low-data target domain. 
Class-conditional GAN (cGAN) extends the standard unconditional GAN to learning data-label distributions and hence capable of generating images belonging to specific classes. Observing that class-conditioning causes mode collapse in limited data settings, where unconditional learning leads to satisfactory generative ability, \cite{shahbazi2022collapse} suggested a transitional training strategy for cGANs that effectively prevents mode-collapse by leveraging unconditional learning. This strategy starts with an unconditional GAN and gradually injects conditional information into the generator and the objective function. 

Diffusion model is a parameterized Markov chain trained with variational inference to produce samples matching the data after finite steps \cite{denoisingDiffusion}. Transitions of this chain are learned to reverse a diffusion process, which is a Markov chain that gradually adds noise to the data in the opposite direction of sampling until the signal is destroyed. Showing that diffusion models can achieve image sample quality superior to the current state-of-the-art GANs, \cite{guidedDiffusion} improved the synthesized image quality with a classifier guidance.

In our proposed data augmentation pipeline, both cGAN with transitional training strategy \cite{shahbazi2022collapse} and the guided-diffusion model \cite{guidedDiffusion} are adopted and experimented.

\section{Method}

In this section, we are going to talk about why we choose SSIM as the basis of our proposed metric in Section \ref{sec:SSIM}, how we group the dataset sub-classes (i.e., the original classes) into super-classes to compute the metric value in Section \ref{sec:supClsInfo}, how the metric is constructed in Section \ref{sec:SSIM-supSubClsMetric}, and how we are going to augment the imbalanced image dataset using our proposed pipeline in Section \ref{sec:GM-augCls}.

\subsection{Structural Similarity}\label{sec:SSIM}

As introduced in \cite{HVS2002}, human visual system is able to assess the perceptual similarity between two images and highly sensitive to structural differences. To measure images differences, naive metrics such as mean squared error (MSE) and peak signal-to-noise ratio (PSNR) cannot capture the perceived visual quality. Hence as an alternative, Structural Similarity Index (SSIM) \cite{SSIM2004} seeks to compare the structures of the reference and the distorted image signals, which indeed correlates more with human cognitive understanding.

The SSIM compares the local patterns of pixel intensities in images that the luminance and contrast have been normalized. It separates the task of similarity measurement into three comparisons: luminance, constrast and structure. Suppose $\boldsymbol{x}$ and $\boldsymbol{y}$ are two images aligned with each other. Denote the mean intensity as $\mu_x$ and $\mu_y$, the standard deviation as $\sigma_x$ and $\sigma_y$, and the covariance as $\sigma_{xy}$. Choose the constant $C_1=(K_1 L)^2$ where $L$ is the dynamic range of the pixel values (255 for 8-bit grayscale images), and $K_1\ll 1$ is a small constant. Similarly, choose constant $C_2=(K_2 L)^2$, and $K_2\ll 1$. Then these three components can be combined to yield a final image similarity measurement:
\begin{equation}
    \mathrm{SSIM}(\boldsymbol{x},\boldsymbol{y})=\frac{(2\mu_x\mu_y + C_1)(2\sigma_{xy} + C_2)}{(\mu_x^2+\mu_y^2+C_1)(\sigma_x^2+\sigma_y^2+C_2)}
\end{equation}

The SSIM metric returns one when the two images $\boldsymbol{x}$ and $\boldsymbol{y}$ are identical and decreases as differences become more relevant. A value of zero indicates no structural similarity. In this paper, we extend the concept of structural similarity to image dataset classes hierarchy as shown in Section \ref{sec:supClsInfo}.

\subsection{Super-class Information for Image Dataset}\label{sec:supClsInfo}

\begin{figure}[h]
\centering
\includegraphics[width=1.0\linewidth]{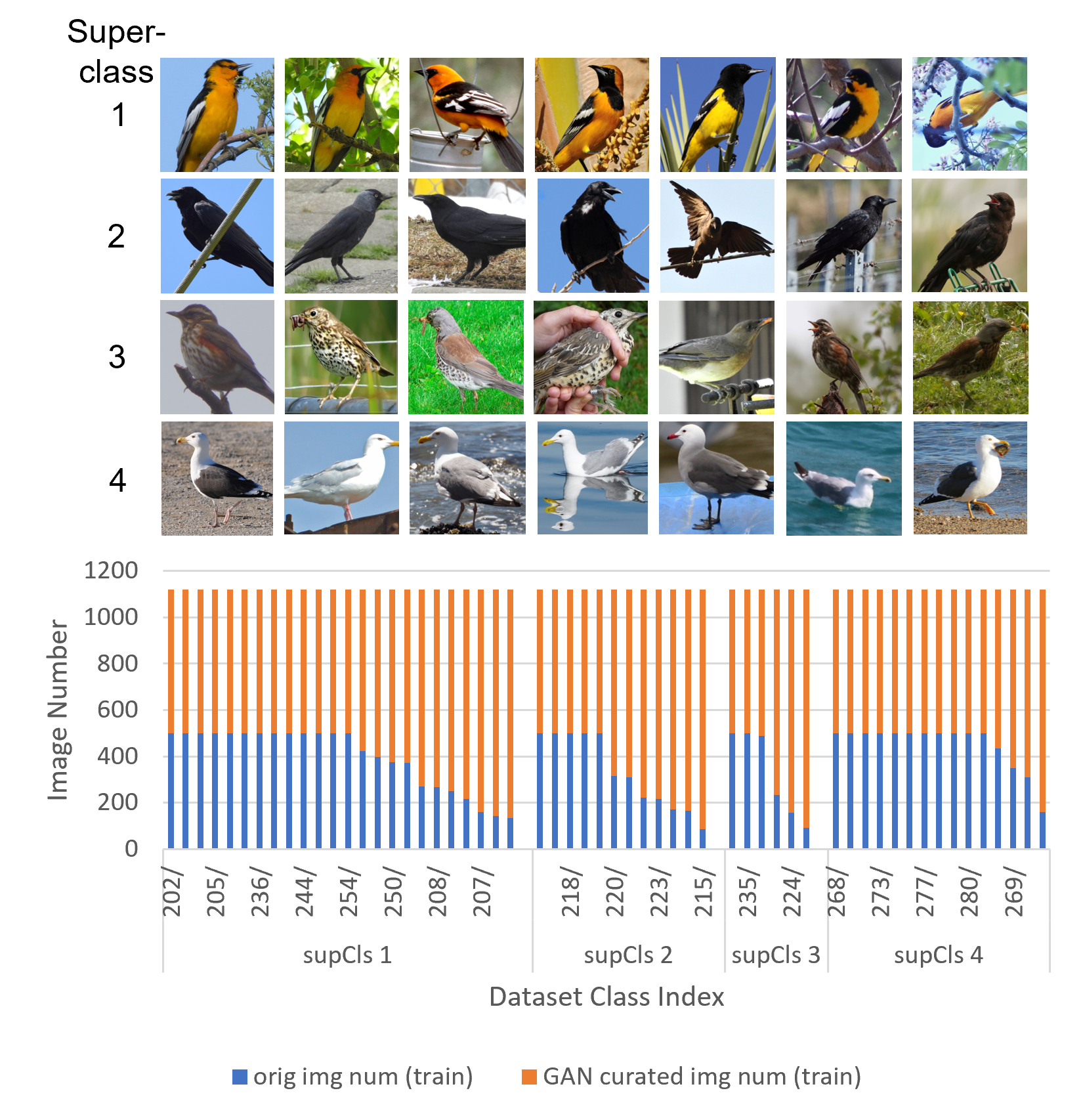}
\caption{({\it Top}) Sample images of the original iNaturalist-2019 Birds dataset from different super-classes (here only 4 of 9 super-classes are presented for visualization). Though images are from different sub-classes, within each super-class they share similar bird attributes and thus are grouped together. ({\it Bottom}) Bar plot for its class distribution. Notice that the distribution of the original dataset is imbalanced and the one after GM-augCls is balanced.}
\label{fig:classDistribSamp}
\end{figure}

Real world datasets usually follow long-tail instead of uniform class distributions. Figure \ref{fig:classDistribSamp} shows the class distribution of iNaturalist-2019 Birds \cite{iNaturalist2019} as an example of such imbalanced dataset.

A naive image dataset similarity measurement would be directly computing the average of SSIM values of every image versus the other images in the dataset. Unfortunately, this will be computationally expensive for large scale datasets. Thus, we seek to alleviate the time complexity by bootstrapping. Basically, what we want to calculate is an expectation of the SSIM within an image set, as in Eq. \ref{eq:ssimMerge}. In practice, each time we can randomly pick a pair of images $\mathbf{I}_i \neq \mathbf{I}_j$ from the set $\mathcal{X}$ and evaluate their SSIM. This procedure should be repeated enough times (here we use $2\times$ number of images in the whole imbalanced dataset) until the average of SSIM values converge, and the final similarity measurement would be the mean of the SSIM values on each pair of images.

\begin{equation}\label{eq:ssimMerge}
    \mathrm{SSIM_{set}}(\mathcal{X})=\mathbb{E}[\underset{\mathbf{I}_i, \mathbf{I}_j \in_{R}\mathcal{X}}{\mathrm{SSIM}}(\mathbf{I}_i, \mathbf{I}_j)]
\end{equation}

If we simply set $\mathcal{X}$ consisting all the images in the whole imbalanced training set, we can get a straightforward metric, called Merged-Class Structural Similarity (SSIM-mergeCls). However, through mixing up images from all classes, we are in the risk of losing important information of the original dataset class structure. Thus, instead of using the flat structure (all classes at same level) for a classification task, we use a hierarchical class structure based on some ontology or human annotations.

\begin{figure*}
\centering
\includegraphics[width=0.95\textwidth]{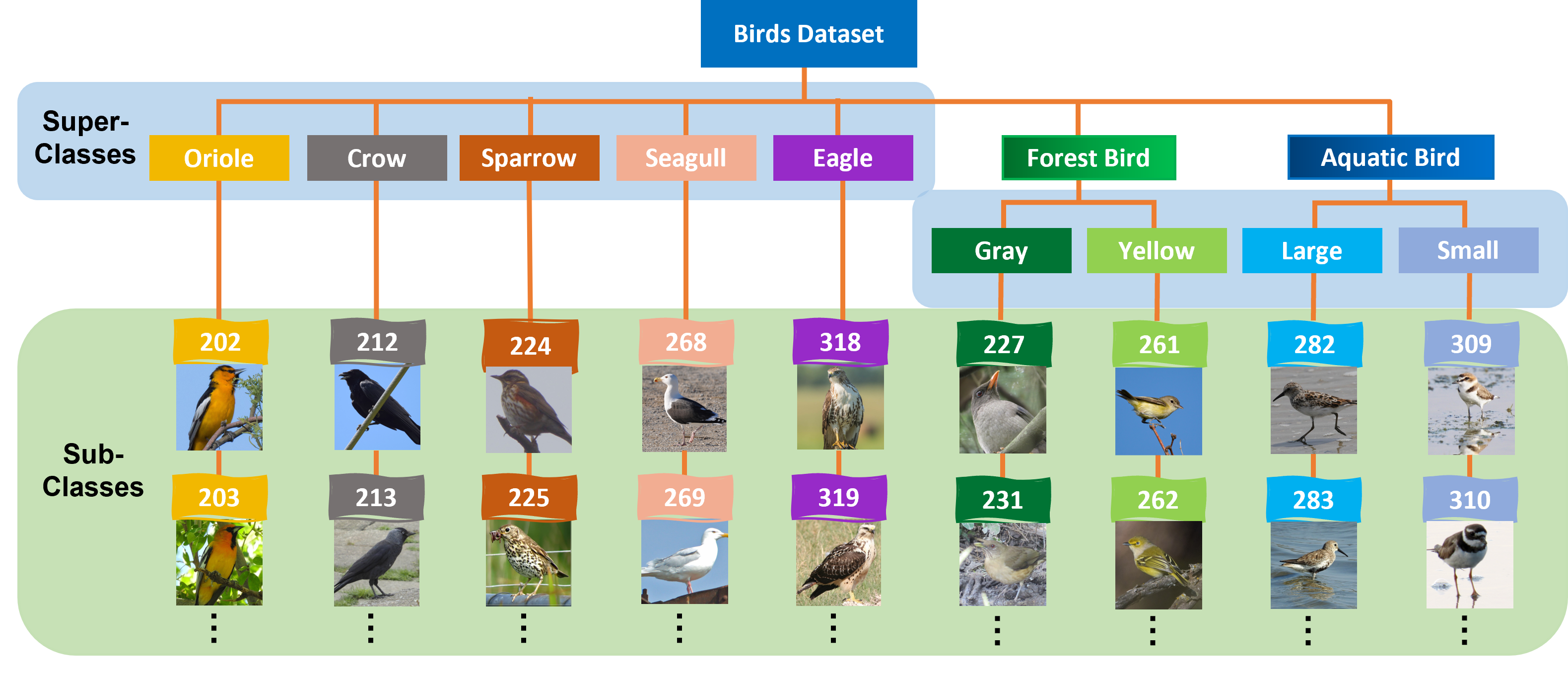}
\caption{The resulting class hierarchy of iNaturalist-Birds dataset by grouping the original sub-classes. The whole dataset contains 9 super-classes and each super-class consists of various sub-classes.}
\label{fig:birds_supClsArch}
\end{figure*}

One may notice that many existing image datasets for classification tasks can be further grouped into broader classes (i.e., super-classes) which subsume several of the original classes (i.e., sub-classes) that keep same image attributes under specific metrics \cite{wen2022learn}. This super-class information can be used in this scenario. For a natural world creatures dataset like iNaturalist-2019 Birds in Figure \ref{fig:classDistribSamp}, these super-classes are formed through human inspections in a top-down manner, as in Figure \ref{fig:birds_supClsArch}: 

\begin{enumerate}
    \item First, all the sub-classes in the original dataset are grouped according to their species. E.g., in Birds dataset, super-class 1 (Oriole), super-class 2 (Crow), super-class 3 (Sparrow), super-class 4 (Seagull) and super-class 5 (Eagle) are grouped in this step.
    \item Then, based on the image background, the sub-classes difficult to be clustered in step 1 are further grouped by their habitats. E.g., forest bird super-class 6 and super-class 7 versus aquatic bird super-class 8 and super-class 9.
    \item Finally, based on the object details, super-classes acquired in step 2 can be further grouped by the bird size, shape and color. E.g., large aquatic bird super-class 8 v.s. small aquatic bird super-class 9.
\end{enumerate}

\begin{table*}
  \centering
  \begin{tabular}{|l|c|c|c|c|c|c|c|c|}
    \hline
    \textbf{Dataset} & UTKFace & Birds & Insects & scene & flowers & Fungi & Reptiles & Amphibians \\
    \hline\hline
    \textbf{sup-cls $^1$} & 5 & 9 & 14 & 6 & 5 & 4 & 2 & 4 \\
    \textbf{sub-cls $^2$} & 5 & 126 & 141 & 6 & 5 & 12 & 39 & 10 \\
    \hline
    \multicolumn{9}{p{300pt}}{$^1$ The number of super-classes in the dataset}\\
    \multicolumn{9}{p{300pt}}{$^2$ The number of sub-classes (i.e., original classes) in the dataset}\\
  \end{tabular}
  \caption{The number of classes in each dataset.}
  \label{table:supClsNum}
\end{table*}

Table \ref{table:supClsNum} provides the number of classes in each dataset we used in this study. Note that for dataset UTKFace, scene and flowers, each super-class only contains one sub-class since the original datasets are already grouped by species. The grouping law for other datasets are provided in Supplemental Material Section S2. With these super-class information, we propose a metric taking a narrower scope of image classes into consideration, called Super-Sub Class Structural Similarity (SSIM-supSubCls). Therefore, a more relevant class similarity measurement is acquired.

\subsection{SSIM-supSubCls Metric}\label{sec:SSIM-supSubClsMetric}

In our proposed SSIM-supSubCls metric, rather than calculating the SSIM from the whole training set $\mathcal{X}$, we compute it from the super-class set $\mathcal{X}_k$, as in Eq. \ref{eq:ssimSupSub}. Practically, each time we randomly select a pair of images $\mathbf{I}_a \neq \mathbf{I}_b$ from the super-class set $\mathcal{X}_k$ and evaluate their SSIM. This process is also repeated enough times (here we use $2\times$ number of images in each super-class set $\mathcal{X}_k$) until the average of SSIM values converge. The similarity measurement on each super-class $k$ would be the mean of the SSIM values on each pair of images in $\mathcal{X}_k$.

Considering the fact that for classification, the classes most difficult to be distinguished are the key impact factors on performance. Thus, our final SSIM-supSubCls measurement would be the maximum of the SSIM values on every super-class set $\mathcal{X}_k$, which indicates the largest similarity among all super-classes.

\begin{equation}\label{eq:ssimSupSub}
    \mathrm{SSIM_{supSubCls}}(\mathcal{X})=\underset{\mathcal{X}_k \subset \mathcal{X}}{\max}\{\mathrm{SSIM_{set}}(\mathcal{X}_k)\}
\end{equation}

This metric provides a prior knowledge on whether the deep generative model data augmentation works on the image data classification by measuring its image similarity within super-class set. Considering that deep generative models like cGAN or diffusion model can faithful generate images from different classes which embrace large distinctions like species, but struggle on reconstructing the very subtle distinctions on the image objects, e.g., if there exists black feathers on the birds' wings or not (as in Figure \ref{fig:classDistribSamp} ({\it Top}) super-class 4 Seagulls). it is not the case that deep generative model data augmentation can improve classification performance on every dataset. Therefore, our proposed metric will give the researcher insights on which data augmentation method should use instead of blindly apply the GM-augCls on any image datasets. From the definition in Eq. \ref{eq:ssimSupSub}, higher SSIM-supSubCls value indicates higher similarity among super-classes in the dataset, harder for the deep generative model to reconstruct the details in the images for distinguishing between sub-classes within this super-class, and thus a less chance to get benefits from the GM-augCls.

\subsection{cGAN-aug Classification Pipeline}\label{sec:GM-augCls}

\begin{figure*}[t!]
\centering
\includegraphics[width=\linewidth]{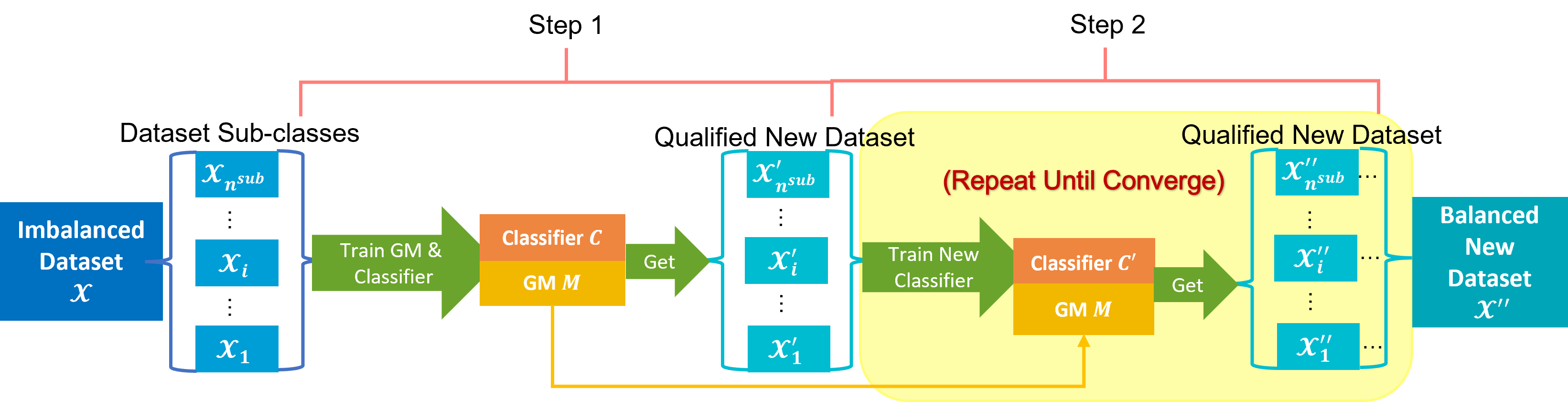}
\caption{The schematic diagram of our proposed Deep Generative Model Data Augmentation Classification (GM-augCls) method. First, a deep generative model (GM) $M$ and a classifier $C(\cdot)$ are trained on the whole imbalanced training set $\mathcal{X}$. $M$ is trained only once, while $C(\cdot)$ gets updated each time by training on the new dataset $\mathcal{X}'$ augmented with qualified new images. This whole procedure is iterated several steps, and in every step $C(\cdot)$ is replaced with $C'(\cdot)$ acquired in the immediate previous step. The augmented $\mathcal{X}''$ is the final balanced dataset we want.}
\label{fig:pipeline_cGANaug}
\end{figure*}

In this section we propose a pipeline to verify that our metric correlates with the deep generative model data augmentation method for improving classification performance. An illustrative figure of the pipeline called GM-augCls is provided in Figure \ref{fig:pipeline_cGANaug}. This figure shows that, in an iterative way, using existing deep generative models like cGAN or diffusion model and classifiers, we can generate a new balanced dataset augmented by \textit{qualified} artificial images. By qualify, we mean, that only synthesized images with large probability during the classification steps are kept to enlarge the dataset.

In practice, we use a StyleGAN2 \cite{karras2020analyzing} based cGAN with transitional training strategy \cite{shahbazi2022collapse} or guided-diffusion model \cite{guidedDiffusion} to generate artificial images as potential ones to augment the imbalanced dataset. These deep generative models are trained on the original imbalanced training set. Then a ResNet \cite{He2015} or Masked Autoencoder (MAE) \cite{MaskedAutoencoders2021} classifier also trained on the original set with sub-class labels are applied on the synthesized images to select qualified ones. The selected images combined with the original ones compose our final balanced dataset. 

Our goal is to enhance the classification performance on an imbalanced image dataset. The deep generative model can generate images belonging to each sub-class under the user's control of synthesized image numbers. However, not all new images fall in a desired sub-class, even with cGAN capable of generating images belonging to specific classes, as the deep generative model is trained with the whole dataset. Thus, we need a classifier to pick images only in the expected class. As shown in Fig \ref{fig:pipeline_cGANaug} step 1, since at the beginning no artificial images have been used to augment the dataset, we use a classifier $C(\cdot)$ trained on the whole original imbalanced set $\mathcal{X}$ to classify each sub-class. Under a prediction probability threshold $\alpha$, $C(\cdot)$ select from all the synthesized images to get \textit{qualified} ones as new members of the augmented balanced set $\mathcal{X}'$. This means that the to-be-balanced dataset and the to-be-improved classifier are combat with each other: on one hand we use the dataset to train the classifier, on the other hand we use the classifier to boost the dataset.  

Thus, the above mentioned process is iterated several steps. In each step the pretrained classifier $C(\cdot)$ is replaced with the $C'(\cdot)$ one trained on the new set $\mathcal{X}'$ acquired in the immediate previous step, until the validation accuracy with $C'(\cdot)$ is not improved under a specific tolerance $\epsilon$. In practice, this is repeated for at most two steps in our experiments. Finally, the augmented set $\mathcal{X}''$ is the balanced dataset we want.

\subsection{Selection on Classifier and Deep Generative Model}\label{sec:cls_gan_selection}

In our work, the classifier used to select qualified images and validate the classification accuracy is ResNet18 \cite{He2015}. To justify our proposed metric and pipeline work regardless of the classifier selection, we also provide the results with Masked Autoencoder (MAE) ViT-H$_{128}$ \cite{MaskedAutoencoders2021} which shows state-of-the-art results in dataset such as \cite{iNaturalist2019}. 

For the GAN architecture, we use StyleGAN2 \cite{karras2020analyzing} based cGAN on the original imbalanced image dataset. Observed that class-conditional learning leads to mode collapse while the unconditional one still maintains satisfactory generative ability under limited data settings, \cite{shahbazi2022collapse} proposed a learning approach which starts with an unconditional GAN and gradually injects class conditional information into the generator and the objective functions. Our pipeline with cGAN adopts this transitional training strategy.

Diffusion models, which can achieve image sample quality superior to the current state-of-the-art GANs, is also experimented in our pipeline. Specifically, guided-diffusion \cite{guidedDiffusion} improving the synthesized image quality with a classifier guidance is used.

\section{Experiments}

\subsection{Experimental Settings}

To evaluate the above proposed methods, we carried out our major experiments on dataset iNaturalist-2019 \cite{iNaturalist2019}, flowers \cite{flowersDatasetKaggle}, UTKFace \cite{UTKFaceDatasetKaggle} and scene \cite{sceneDatasetKaggle} which are either long-tail or imbalanced distributed. The iNaturalist-2019 has around 265K images in the long-tail training set and 3k images in a balanced validation set with 6 categories, and each category contains different number of sub-classes. Here we are using the categories of Insects, Birds, Amphibians, Fungi and Reptiles for experiments. The flowers dataset contains nearly 4k images in the imbalanced training dataset with 5 sub-classes. The UTKFace dataset is a large-scale human face dataset with 5 different ethnicity across long age span. In this work we are predicting their ethnicity. The imbalanced scene dataset contains about 25k images from a wide range of natural scenes with 6 different sub-classes. Class number distributions of all these datasets are shown in Supplemental Material Section S3.

In our experiments, the deep generative models, ResNet18 and MAE classifiers are first trained on the original imbalanced set with sub-class instead of super-class labels. cGAN models are trained until the Frechet Inception Distance (FID) scores converge. The diffusion models are trained for 100k iterations resumed on an ImageNet \cite{deng2009imagenet} pretrained model. To guide the diffusion model during sampling, an additional classifier is also trained until its validation accuracy converges \cite{guidedDiffusion}. The ResNet18 classifiers are trained for 100 epochs and the MAE for 50 epochs when their validation accuracy converges, with their hyper-parameters remaining the same throughout the whole procedure in each case. 

In this paper, we only report the classification results at the final step. The maximum top1 validation accuracy are provided before and after the GM-augCls pipeline. To verify if the SSIM-subSupCls correlates with the classification accuracy improvement, the metric values are measured on the original imbalanced training sets.










\subsection{Qualitative Evaluation Results}

\begin{figure*}[t]
\centering
\includegraphics[width=\linewidth]{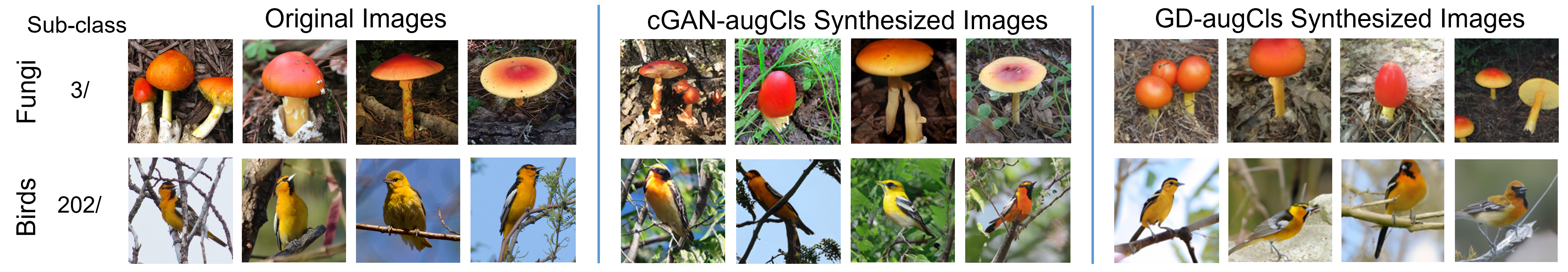}
\caption{Results for qualitative evaluation of our proposed GM-augCls pipeline. These sample images indicate that our pipeline can synthesize high quality images belonging to each sub-class even though the deep generative models are trained on the whole original dataset. More sample images are provided in Supplemental Material Section S4.}
\label{fig:qualiExamples_short}
\end{figure*}

Table \ref{table:FID_SSIMaug} provides the FID scores for the StyleGAN2 based cGAN and GD models. Figure \ref{fig:qualiExamples_short} also provides sample images as a comparison between the original and GM-augCls synthesized ones. They demonstrate our data augmentation method can generate high quality images for each sub-class after classifier selection.

\begin{table}[h!]
\centering
\begin{tabular}{|l|c|c|}
\hline
\textbf{Dataset} & \textbf{StyleGAN2 FID} & \textbf{GD FID} \\
\hline\hline
UTKFace & 4.37 & - \\
Birds & 4.68 & 28.25 \\
Insects & 6.56 & - \\
scene & 9.84 & 54.34 \\
flowers & 15.21 & - \\
Fungi & 8.67 & 50.08 \\
Reptiles & 6.60 & - \\
Amphibians & 11.74 & - \\
\hline
\end{tabular}
\caption{FID scores for StyleGAN2 based cGAN and Guided Diffusion on each dataset.}
\centering
\vspace{-0.8em}
\label{table:FID_SSIMaug}
\end{table}

\subsection{Quantitative Evaluation Results}

We tested our metric and pipeline on dataset iNaturalist-2019, flowers, UTKFace and scene with the original balanced validation set. Table \ref{table:ResNet18_MAE_cls_results} shows their top1 validation accuracy with cGAN and guided-diffusion (GD) models trained on the original and cGAN-augCls or GD-augCls augmented dataset. For ablation study, results on the cGAN augmented datasets \textit{without} image selection are also provided as baseline.

As in Table \ref{table:ResNet18_MAE_cls_results}, for dataset flowers, iNaturalist-2019 Fungi, Reptiles and Amphibians, compared with the ResNet18 classifier trained on the original imbalanced set, the ones trained on the cGAN-augCls balanced set have significantly enhanced accuracy. To certify that this conclusion can be drawn regardless of the classifier selection, we also conduct the experiments with MAE classifier and the same results are obtained. Supplemental Material Section S5 also provides learning curves during this procedure.

\begin{table*}[h!]
    \centering
    \begin{tabular} {|l|c|c|c|c|c|c|c|}
        \hline
        \multirow{2}{*} {\textbf{Dataset}} & \multicolumn{2}{c|}{\textbf{Original Acc}} & \multicolumn{2}{c|}{\textbf{cGAN-aug Acc $^1$}} & \multicolumn{2}{c|}{\textbf{cGAN-augCls Acc}} & {\textbf{GD-augCls Acc}} \\
        \cline{2-8}
        & \textbf{ResNet18} & \textbf{MAE} & \textbf{ResNet18} & \textbf{MAE} & \textbf{ResNet18} & \textbf{MAE} & \textbf{ResNet18} \\
        \hline
        \hline
        UTKFace & 82.31\% & \textbf{87.63\%} & 83.32\% & 86.62\% & \textbf{84.29\%} & 86.82\% & - \\
        Birds & 52.12\% & \textbf{80.42\%} & 52.12\% & 80.16\% & 54.50\% & 80.16\% & \textbf{55.03\%} \\
        Insects & 61.23\% & \textbf{87.23\%} & 59.34\% & 86.29\% & \textbf{61.70\%} & \textbf{87.23\%} & - \\
        scene & 90.12\% & 94.29\% & 90.53\% & 94.35\% & \textbf{90.65\%} & \textbf{94.41\%} & 90.24\% \\
        flowers & 64.14\% & 95.40\% & 79.86\% & 95.86\% & \textbf{80.00\%} & \textbf{96.55\%} & - \\
        Fungi & 50.00\% & \textbf{88.89\%} & 66.67\% & \textbf{88.89\%} & \textbf{75.00\%} & \textbf{88.89\%} & 69.44\% \\
        Reptiles & 28.21\% & 81.20\% & 38.63\% & 79.49\% & \textbf{41.03\%} & \textbf{82.05\%} & - \\
        Amphibians & 20.00\% & 74.67\% & \textbf{43.33\%} & 76.67\% & 36.67\% & \textbf{80.00\%} & - \\
        \hline
        \multicolumn{7}{p{375pt}}{$^1$ When trained on cGAN augmented balanced dataset without classifier selctions (as baseline)}\\
    \end{tabular}
    \caption{Quantitative results of our proposed pipeline with ResNet18 and Masked Autoencoder (MAE) for image selection and evaluation. Here we list the maximum top1 classification validation accuracy within 100 epochs for ResNet18 and 50 epochs for MAE. All classifiers are trained on the original imbalanced set and the balanced set after data augmentation pipeline respectively.}
    \label{table:ResNet18_MAE_cls_results}
\end{table*}

As a well-known regularization-based method, class-balanced loss (CB-loss) \cite{cui2019classbalancedloss} is specifically designed to add weights for each class in the loss function. For comparison, Table \ref{table:CBloss_results} also provides the top1 validation accuracy training with ResNet18 CB-loss on the original dataset. Compared with the results in Table \ref{table:ResNet18_MAE_cls_results}, using Effective Number Class Balanced loss does not improve the accuracy as much as our proposed pipeline.

\begin{table*}[h!]
\centering
\begin{tabular}{|l|c|c|c|c|}
\hline
\textbf{Dataset} & UTKFace & scene & flowers & Amphibians \\
\hline\hline
\textbf{CB-loss Acc} & 73.65\% & 87.71\% & 75.86\% & 20.00\% \\
\hline
\end{tabular}
\caption{Top1 validation accuracy with ResNet18 CB-loss training on the original dataset.}
\centering
\vspace{-0.8em}
\label{table:CBloss_results}
\end{table*}

To verify that our proposed metric correlates with the accuracy improvement, we measure the SSIM-mergeCls and SSIM-supSubCls values for each dataset. Table \ref{table:SSIM_value_results} summarises their metric values. This table shows that the deep generative model data augmentation method only works when the metric values are small. This observation confirms our hypothesis that for an image dataset with high similarity between classes, the deep generative model suffers in unfaithfully reconstructing the subtle distinctions between class features and the classifiers struggles on distinguishing various classes, which leads to a lower classification accuracy improvement.


\begin{table*}[h!]
    \centering
    \begin{tabular} {|l|c|c|c|c|}
        \hline
        \textbf{Dataset} & \textbf{SSIM-mergeCls} & \textbf{SSIM-supSubCls} & \textbf{cGAN-augCls Acc Impr $^1$} & \textbf{cGAN-augCls Works}\\
        \hline
        \hline
        UTKFace & 0.3649 & 0.3834 & 2.41\% & $\times$ \\
        Birds & 0.1698 & 0.3742 & 4.57\% & $\times$ \\
        Insects & 0.1373 & 0.3115 & 0.77\% & $\times$ \\
        scene & 0.1462 & 0.2625 & 0.59\% & $\times$ \\
        flowers & 0.1319 & 0.1652 & 24.73\% & $\checkmark$ \\
        Fungi & 0.0617 & 0.0880 & 50.00\% & $\checkmark$ \\
        Reptiles & 0.0645 & 0.0793 & 45.44\% & $\checkmark$ \\
        Amphibians & 0.0725 & 0.0792 & 83.35\% & $\checkmark$ \\
        \hline
        \multicolumn{5}{p{350pt}}{$^1$ The relative accuracy improvement on cGAN-augCls Acc with ResNet18 classifier (calculated from Table \ref{table:ResNet18_MAE_cls_results} normalized by the original accuracy)}\\
    \end{tabular}
    \caption{This table provides the SSIM-mergeCls and SSIM-supSubCls metric values for each dataset. Based on the relative accuracy improvement, whether the cGAN-augCls pipeline works or not is also denoted. Note that the cGAN data augmentation only works when the SSIM-mergeCls values are smaller than threshold 0.1319, or for the SSIM-supSubCls smaller than threshold 0.1652.
    }
    \label{table:SSIM_value_results}
\end{table*}

\begin{figure}[h]
\centering
\includegraphics[width=1.0\linewidth]{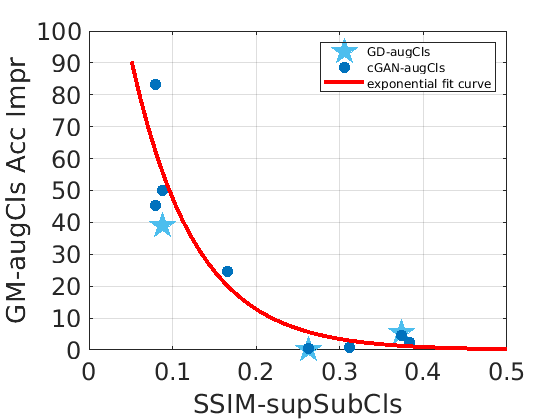}
\caption{Scatter plot and curve fitted for dataset SSIM-supSubCls values vs GM-augCls accuracy improvements for class-conditional StyleGAN2 and Guided Diffusion model trained on iNaturalist-2019, flowers, UTKFace and scene (calculated from Table \ref{table:SSIM_value_results}). Regardless of model architecture or dataset, the results show a common exponential decay trend.}
\label{fig:table3FittedCurveWithDiffu}
\end{figure}

We further quantify the relationship between SSIM-supSubCls values and GM-augCls relative accuracy improvement in Table \ref{table:SSIM_value_results}. As shown in Figure \ref{fig:table3FittedCurveWithDiffu}, the top1 validation accuracy increment decays exponentially with respect to the metric values, and this exponential function can be fitted as:

\begin{equation}\label{eq:exp-decay}
    \hat{f}(x) = 0.94^{202.74x-79.92}
\end{equation}

With this exponential curve fitted, one can easily predict the expected accuracy improvement for any image dataset from its SSIM-supSubCls value. Table \ref{table:rSquare-MAE-fit} provides goodness-of-fit ($R^2$) and Mean Absolute Error (MAE) measurements for the function $f$. The high $R^2$ and low MAE values show that the formulation of $f$ is highly effective on modeling the relationship between SSIM-supSubCls and accuracy improvement with our proposed data augmentation pipeline.

\begin{table}[t]
\begin{center}
\begin{tabular}{|l|c|c|}
\hline
\textbf{Model} & \textbf{$R^2_{f}$} & \textbf{MAE$_{f}$(\%)$^1$} \\
\hline\hline
cGAN-augCls & 0.8749 & 4.7882 \\
GD-augCls & 0.9597 & 5.5035 \\
\hline
\multicolumn{3}{p{210pt}}{$^1$ Here MAE is short for Mean Absolute Error}
\end{tabular}
\end{center}
\vspace{-0.8em}
\caption{Goodness-of-fit $R^2$ and Mean Absolute Error (MAE) measurements for the SSIM-supSubCls vs GM-augCls function $f$.}
\label{table:rSquare-MAE-fit}
\end{table}

Table \ref{table:compareWithSOTA_results} compares the state-of-the-art (SOTA) classification validation accuracy from \cite{MaskedAutoencoders2021} and ours on the whole iNaturalist-2019 dataset. From this table, our results with  Masked Autoencoder (MAE) is comparable with the SOTA one, even though the SOTA is with $448\times 448$ while ours is with $128\times 128$ input image resolutions which largely reduce the need of running time and computational resources. 

\begin{table}[h!]
    \centering
    \begin{tabular} {|l|c|c|c|c|}
        \hline
        \multirow{2}{*} {\textbf{SOTA $^1$}} & \multicolumn{2}{c|}{\textbf{Original Acc}} & \multicolumn{2}{c|}{\textbf{cGAN-augCls Acc}} \\
        \cline{2-5}
        & \textbf{ResNet18} & \textbf{MAE} & \textbf{ResNet18} & \textbf{MAE}\\
        \hline
        \hline
        \textbf{88.3\%} & 54.4\% & \textbf{83.1\%} & 55.4\% & 82.5\% \\
        \hline
        \multicolumn{5}{p{210pt}}{$^1$ State-of-the-art (SOTA) classification validation accuracy with Masked Autoencoder ViT-H$_{448}$ \cite{MaskedAutoencoders2021}}\\
    \end{tabular}
    \caption{Top1 classification validation accuracy compared the SOTA and ours on the whole iNaturalist-2019 dataset.}
    \label{table:compareWithSOTA_results}
\end{table}



\section{Conclusion and Discussion}

We propose a metric called Super-Sub Class Structural Similarity (SSIM-supSubCls) to evaluate the image set class similarity serving as an indicator of classification accuracy improvement after data re-balancing with deep generative model. We also provide a new pipeline achieving data augmentation efficiently for imbalanced image datasets, using cGAN or diffusion models and ResNet or Masked Autoencoder (MAE) classifiers. With user controlling the number of images generated for each sub-class, the original dataset can be re-balanced with qualified images.

We applied our pipeline on imbalanced dataset iNaturalist-2019, flowers, UTKFace and scene, evaluating their classification accuracy when trained on the original and balanced dataset after augmentation. Experimental results reveal that, the top1 validation accuracy increment decays exponentially with respect to the SSIM-supSubCls values regardless of model architecture or dataset. With this exponential curve fitted, one can easily predict the expected accuracy improvement for any image dataset from its metric value.



\end{document}